\documentclass{ptephy_v2}%%%%where ptephy_v1 is the template name

\preprintnumber{2502-008} %%% %%% Insert preprint number here
\usepackage{hyperref}
\usepackage{siunitx}

\usepackage{hyperref}
\usepackage{subfiles}
\usepackage{xcolor}
\usepackage{lineno}

\renewcommand{\textbf}[1]{{\bfseries\sffamily#1}} %太字フォントの決定

\newcommand{\figrefin}[1]{Fig.~\ref{#1}}%図表等番号のラベル省略
\newcommand{\figref}[1]{Figure~\ref{#1}}%図表等番号のラベル省略
\newcommand{\tabref}[1]{Table~\ref{#1}}
\newcommand{\secref}[1]{Section~\ref{#1}}

\newcommand{\refref}[1]{Ref.~\cite{#1}}

\newcommand{\vrange}[2]{\ensuremath{#1\text{--}#2}}
\newcommand{\vals}[2]{\ensuremath{#1\pm #2}}
\newcommand{\RI}[2]{\ensuremath{{}^{#1}\mathrm{#2}}}
\newcommand{\RIU}[2]{\ensuremath{{}^{#1}\mathrm{#2}}(U)}
\newcommand{\RITh}[2]{\ensuremath{{}^{#1}\mathrm{#2}}(Th)}
\newcommand{\betagamma}{\ensuremath{\mathrm{\beta/\gamma}}\,}
\newcommand{\alpharay}{$\mathrm{\alpha}$-ray\,}

\newcommand{\keVee}{\ensuremath{\mathrm{keV_{ee}}}}
\newcommand{\dru}{\ensuremath{\mathrm{d^{-1} kg^{-1} keV^{-1}}}}

\newcommand{\druasmall}{\ensuremath{\mathrm{events/(d\cdot kg\cdot keV)}}}
\newcommand{\activitykg}{\si{\micro \becquerel/ \kilogram}}
\newcommand{\genergy}[1]{\ensuremath{E_\mathrm{\gamma}=#1\ \si{keV}}}
\newcommand{\qbeta}[1]{\ensuremath{Q_\mathrm{\beta}=#1\ \si{MeV}}}
\newcommand{\qec}[1]{\ensuremath{Q_\mathrm{EC}=#1\ \si{MeV}}}
\newcommand{\halflive}[2]{\ensuremath{T_\mathrm{1/2}=#1\ \si{#2}}}

\newcommand{\picolonlong}{Pure Inorganic Crystal Observatory for LOw-energy Neutr(al)ino}

\newcommand{\prevterm}{2021 Jul. 12--2021 Aug. 23 (JST)}
\newcommand{\thisterm}{2021 Nov. 11--2022 May 24 (JST)}
\newcommand{\DMupperlimitVal}{4.3}
\newcommand{\DMupperlimit}{$S_\mathrm{m}=\DMupperlimitVal $~\dru}

\newcommand{\livetime}{\SI{156.0}{d}}
\newcommand{\exposure}{\livetime$\times$\SI{1.34}{\kilogram}\,}
\newcommand{\Qprehit}{\ensuremath{Q_\text{pre--hit}\,}}
\newcommand{\roirangeval}{2--6}
\newcommand{\roirange}{\roirangeval~\keVee\,}

\begin{document}

\title{Background stability and annual modulation test using PICOLON ultra-pure NaI(Tl) detector}

\author{K.~Kotera$^{*}$}
\affil{Graduate School of Advanced Technology and Science for Innovation, Tokushima University, 2-1 Minami Josanjimacho, Tokushima, Tokushima, 770-8506, Japan \email{c612344002@tokushima-u.ac.jp}\collaborator{PICOLON Collaboration}}

\author{D.~Chernyak}
% \affil{Department of Physics and Astronomy, University of Alabama, Tuscaloosa, AL 35487, USAJapan}
\affil{Research Center for Neutrino Science, Tohoku University, 6-3 Aramaki-aza-aoba, Aobaku, Sendai, Miyagi, 980-8578, Japan}

\author{H.~Ejiri}
\affil{Research Center for Nuclear Physics, Osaka University, 10-1 Mihogaoka, Ibaraki, Osaka, 567-0047,}

\author{K.~Fushimi}
\affil{Department of Physics, Tokushima University, Tokushima University, 2-1 Minami Josanjimacho, Tokushima, Tokushima, 770-8506, Japan}

\author[2]{K.~Hata}
% \affil{Research Center for Neutrino Science, Tohoku University, 6-3AramakiAzaAoba, Aobaku, Sendai, Miyagi, 980-8578, Japan}

\author{R.~Hazama}
\affil{Department of Environmental Science and Technology, Osaka Sangyo University, 3-1-1 Nakagaito, Daito, Osaka, 574-8530, Japan}

\author{T.~Iida}
\affil{Faculty of Pure and Applied Sciences, University of Tsukuba, 1-1-1 Tennoudai, Tsukuba, Ibaraki, 305-8571, Japan}

\author[2]{H.~Ikeda}

\author{K.~Imagawa}
\affil{I.S.C.Lab, 5-15-24 Torikai Honmachi, Settsu, Osaka, 566-0052, Japan}

\author[2,8]{K.~Inoue}
\affil{Kavli Institute for the Physics and Mathematics of the Universe (WPI), 5-1-5 Kashiwanoha, Kashiwa, Chiba, 277-8583, Japan}

\author{H.~Ito}
% \affil{Department of Physics, Faculty of Science and Technology, Tokyo University of Science, Noda, Chiba 278-8510, Japan}
\affil{Department of Physics, Kobe University, Rokkodai-cho, Hyogo 657-8501, Japan}

\author[3]{T.~Kishimoto}
% \affil{Department of Physics, Osaka University, 1-1 Machikaneyama-cho, Toyonaka, Osaka, 560-0043, Japan}

\author[2,8]{M.~Koga}

\author{A.~Kozlov}
\affil{National Research Nuclear University ”MEPhI” (Moscow Engineering Physics Institute), Moscow, 115409, Russia}

\author[8,11]{K.~Nakamura}
\affil{Butsuryo College of Osaka, 3-33 Ohtori Kitamachi, Nishi ward, Sakai, Osaka, 593-8328, Japan}

\author[4]{R.~Orito}
\author[3]{T.~Shima}
\author[8,12]{Y.~Takemoto}
\affil{Kamioka Observatory, Institute Cosmic Ray Research, University of Tokyo, Kamioka, Gifu, 506-1205, Japan}

\author[3]{S.~Umehara}
\author{Y.~Urano}
\affil{Institute for Materials Research, Tohoku University, 2-1-1 Katahira, Aobaku, Sendai, Miyagi, 980-8577, Japan}

\author[7]{K.~Yasuda}
\author{S.~Yoshida}
\affil{Department of Physics, Osaka University, Toyonaka, Osaka, 560-0043, Japan}

\begin{abstract}%
The dark matter observation claimed by the DAMA/LIBRA experiment has been a long-standing puzzle within the particle physics community.
NaI(Tl) crystals with radiopurity comparable to DAMA/LIBRA's are essential for adequate verification.
Existing experiments using NaI(Tl) target have been hampered by the high radioactivity concentration of NaI(Tl) crystals.
PICOLON experiment conducts an independent search for Weakly Interacting Massive Particles using highest purity NaI(Tl) crystals.
In 2020, the NaI(Tl) crystal (Ingot\#85) reached the same purity level as DAMA/LIBRA crystals.

The DAMA/LIBRA group has stressed that verifying their signal requires high-purity NaI(Tl) crystals with long-term stability.
Based on a six-month measurement, we have confirmed the long-term stability of its radiopurity. 
This stability provides a significant advantage for future efforts to adequately verify the DAMA/LIBRA result using NaI(Tl) crystal. 
In this paper, we present the background stability of purity in the Ingot\#94 NaI(Tl) detector, which was produced using the Ingot\#85 purification method, along with the first annual modulation search conducted by the PICOLON experiment.
\end{abstract}

\subjectindex{F40, F41}
% \linenumbers % Enable line No.

\maketitle

\section{Introduction}
Dark matter (DM) is one of the most important candidates in cosmology and particle physics. Observations of the cosmic microwave background show that DM accounts for 26.4\% of the total energy density of the universe \cite{refId0A,refId0B}.
Theoretical models beyond the standard model are necessary to explain it.
Direct detection of DM is expected to accelerate progress in both experimental astro physics and theoretical studies.
 Weakly Interacting Massive Particles (WIMPs) remains one of the most important candidates for DM.
Recently, lighter WIMPs of sub-GeV mass have been proposed~\cite{10.1093/ptep/ptac097}. Lower-background detectors are crucial for WIMP searches across all mass scales.

An annual modulation signal reported by the DAMA/LIBRA experiment has not been confirmed by other groups operating highest purity NaI(Tl) detectors \cite{10.21468/SciPostPhysProc.12.025}.
At present, the results from dark matter search experiments using xenon targets (XENONnT~\cite{xenon2023dark}, LUX-ZEPLIN~\cite{PhysRevLett.131.041002}, PandaX-4T~\cite{PhysRevLett.127.261802}), and germanium targets (SuperCDMS~\cite{PhysRevLett.120.061802}) have placed strong upper limits on the DAMA/LIBRA signal.
Even the DAMA/LIBRA experiment, which reported the highest sensitivity with NaI(Tl) detectors, has a background (BG) level of approximately six orders of magnitude higher than the XENONnT experiment.
Although the difference in target nuclei explains part of the discrepancy, it does not fully account for the amplitude or phase of the annual modulation reported in DAMA/LIBRA experiment.
NaI-based experiments such as ANAIS~\cite{ANAIS2021Modulation} and COSINE~\cite{Adhikari2023} have also reported results that disfavor the annual modulation observed by DAMA/LIBRA.
These experiments show time-dependent variations in the BG level, complicating the interpretation of potential modulations.
As discussed in \refref{10.21468/SciPostPhysProc.12.025}, to verify the DAMA/LIBRA experiment, it is adequately required that NaI(Tl) crystals and experimental setups are required to be comparable properties. 
The DAMA/LIBRA NaI(Tl) crystals do not show a significant decrease in event rate with time. 
The development of the purest NaI(Tl) crystal suggests the possibility of a stable-BG.

\begin{table}[!h]
\caption{Concentration of RIs and BG rate reported by DM experiments using NaI(Tl) crystals. The unit of dru stands for \druasmall.}
\label{Tab:RIConcentrationGroups}
 \centering
 \scalebox{0.85}
 {
  \begin{tabular}{ccccc|ccc} \hline
   RIs & DAMA/LIBRA & COSINE & ANAIS-112 & SABRE& Ingot\#85 & Ingot\#94 & Our goal\\
    & \cite{BERNABEI2008297} & \cite{COSINE2020Clystals} & \cite{ANAIS2021Modulation,ANAIS2019Clystal}& \cite{SABRE2205,SABRENaITl2022} & \cite{fushimi2021development} & (30 d)\cite{Kotera2023picolon} & \\\hline
  $\mathrm{^{40}}$K [\activitykg] & $<600$    & $<1060$   & \vrange{545}{1200} & $120$     & --    & -- & $600$ \\
  $^{232}$Th [\activitykg]     & \vrange{2}{31}  & \vrange{2.5}{35} & $4$     & $0.8$     & \vals{1.2}{1.4} & \vals{4.6}{1.2} & $10$  \\
  $^{226}$Ra [\activitykg]     & \vrange{8.7}{124} & \vrange{11}{451} & $10$    & $5$      & \vals{13}{4} & \vals{7.9}{4.4}& $10$  \\
  $^{210}$Pb [\activitykg]     & \vrange{5}{30}  & \vrange{10}{3000} & \vrange{740}{3150}  & $360$     & $<$5.7 & \vals{19}{6} & $50$  \\
  BG rate (dru)           & 1       & 3      & 3.605$\pm$0.003 & 1.39$\pm$0.02 & -- & -- & 1 \\\hline
  \end{tabular}
 }
\end{table}

PICOLON (\picolonlong) is a DM search experiment using ultra-pure NaI(Tl) detectors.
One of PICOLON experiment aims to investigate the annual modulation signal reported by the DAMA/LIBRA experiment.
The main BG sources in NaI(Tl) crystals are \RI{40}{K} (\RI{\mathrm{nat}}{K}), \RI{232}{Th} (Th-series), \RI{226}{Ra} and \RI{210}{Pb} (U-series). In particular, \RI{40}{K} and \RI{210}{Pb} are the main BG contributors at energies below 100~\keVee, where \keVee\ is an observed energy that is calibrated by electron energy.
\tabref{Tab:RIConcentrationGroups} shows the concentrations of radioactive impurities (RIs) in the NaI(Tl) crystals used by DAMA/LIBRA, COSINE, ANAIS-112, SABRE, and PICOLON experiments.
In 2020, we developed an ultra-pure NaI(Tl) crystal named Ingot\#85 using an optimized purification method that includes recrystallization and resins \cite{fushimi2021development}.
According to \tabref{Tab:RIConcentrationGroups}, NaI(Tl) crystals of equal or better purity than DAMA/LIBRA crystals have not been developed, except for the PICOLON experiment.
Thus, the PICOLON ultra-pure NaI(Tl) crystals provide the only viable opportunity to adequately verifying the DAMA/LIBRA experiment.
The Ingot\#94 crystal was produced following the same purification method as Ingot\#85~\cite{fushimi2021development}. 
An initial one-month analysis confirmed that its radiopurity was comparable to that of Ingot\#85~\cite{Kotera2023picolon}. 
This result demonstrated the reproducibility of the purification method.
In this study, we investigated the stability of purity for the NaI(Tl) crystal, Ingot\#94, using a new long-term dataset.
The experimental setup and data acquisition system (DAQ) are described in \secref{sec:experimental_setup}. 
The currently BG result is presented in \secref{sec:alpharesult}, \ref{sec:betagammaspect}. A noise reduction method are explained in \ref{sec:noisereduction}. 
\secref{sec:Limit} presents the test of annual modulation and the DM limits on WIMPs.
\section{Experimental setup}\label{sec:experimental_setup}
\begin{figure}[!h]
 \centering
 \includegraphics[width=0.5\hsize]{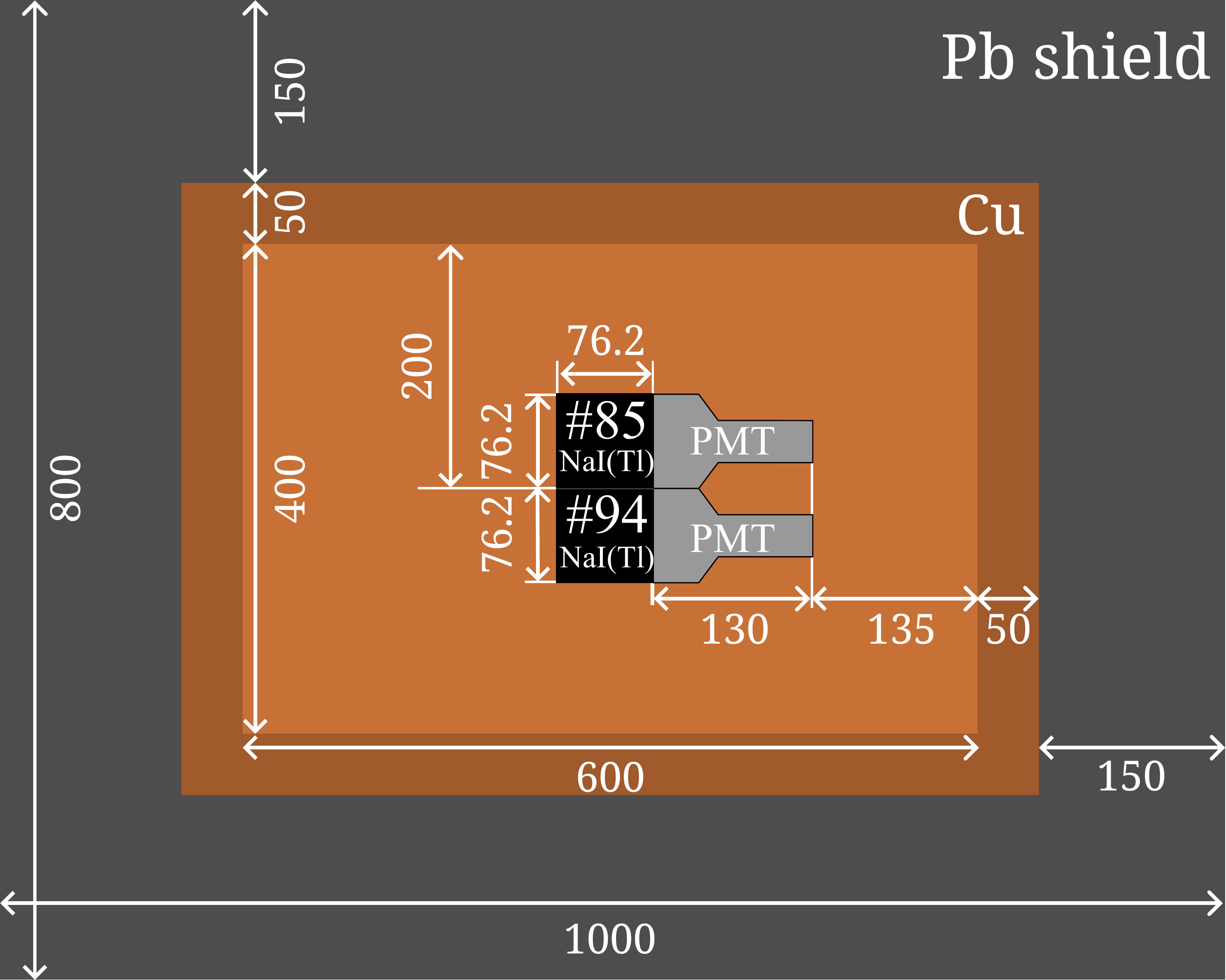}
 \caption{Schematic view of the PICOLON shield (length unit: \si{mm}).}
 \label{fig:NaIdetectorAndShield}
\end{figure}
The PICOLON experimental room is located at the KamLAND area in Kamioka Underground Laboratory at a depth of 2700 meters water equivalent (m.w.e).
\figref{fig:NaIdetectorAndShield} shows a schematic drawing of NaI(Tl) detectors in PICOLON shield.
Ingots \#85 and \#94 were installed into the shielding made of 5~cm thick oxygen-free copper blocks and 15~cm thick low-activity lead blocks.
The radon concentration in experimental room air is 20--50~\si{\becquerel/\meter^3}, and the shield was filled with pure boiled-off nitrogen gas to purge radon.

We assembled two detector modules using NaI(Tl) crystals Ingot\#85 and \#94.
Ingot\#94 was produced using the same purification method and detector design as Ingot\#85, and shows comparable performance in volume and light yield.  
Ingot\#85 was used for continuous data acquisition and anti-coincidence testing.  
However, it was excluded from the present analysis, which focuses on the radiostability of the ultra-pure NaI(Tl) crystal and the annual modulation study of Ingot\#94.

The cylindrical crystals have a diameter of~76.2~mm and a length of~76.2~mm.
The bottom and side surfaces of the crystals were covered with Enhanced Specular Reflector Films (ESR\texttrademark) to guide scintillation photons into an optical window. % $^{\mathrm{TM}}$
The crystal was encapsulated into an acrylic housing and sealed with a 1~cm-thick optical window of synthetic quartz.
The assembled NaI(Tl) crystal was read using a 3-inch low-background Hamamatsu R11065-20 photomultiplier tube (PMT).
Ingot\#94 detector has light yields of about 10~\si{photoelectrons/\kilo\electronvolt}.
The PICOLON data acquisition (DAQ) system consists of a simple trigger from two PMT signals.
Each detector logic signal is generated by a discriminator using an amplified NaI(Tl) linear signal. Detector logic signals issue the 2~\si{\micro s} long OR gate trigger.
NaI(Tl) signals were recorded by MoGURA Flash-ADC board (see \refref{MoGURA}). 

In this study, we analyzed data from a new measurement period, \thisterm, which is independent of the previous dataset for \prevterm~\cite{Kotera2023picolon}.
The exposure time was \exposure\ (realtime: 192.6~\si{d}), from \thisterm.

\section{Radioactivity in NaI(Tl) crystal} \label{sec:alpharesult}
\begin{figure}[!h]
  \centering
  \includegraphics[width=0.65\hsize]{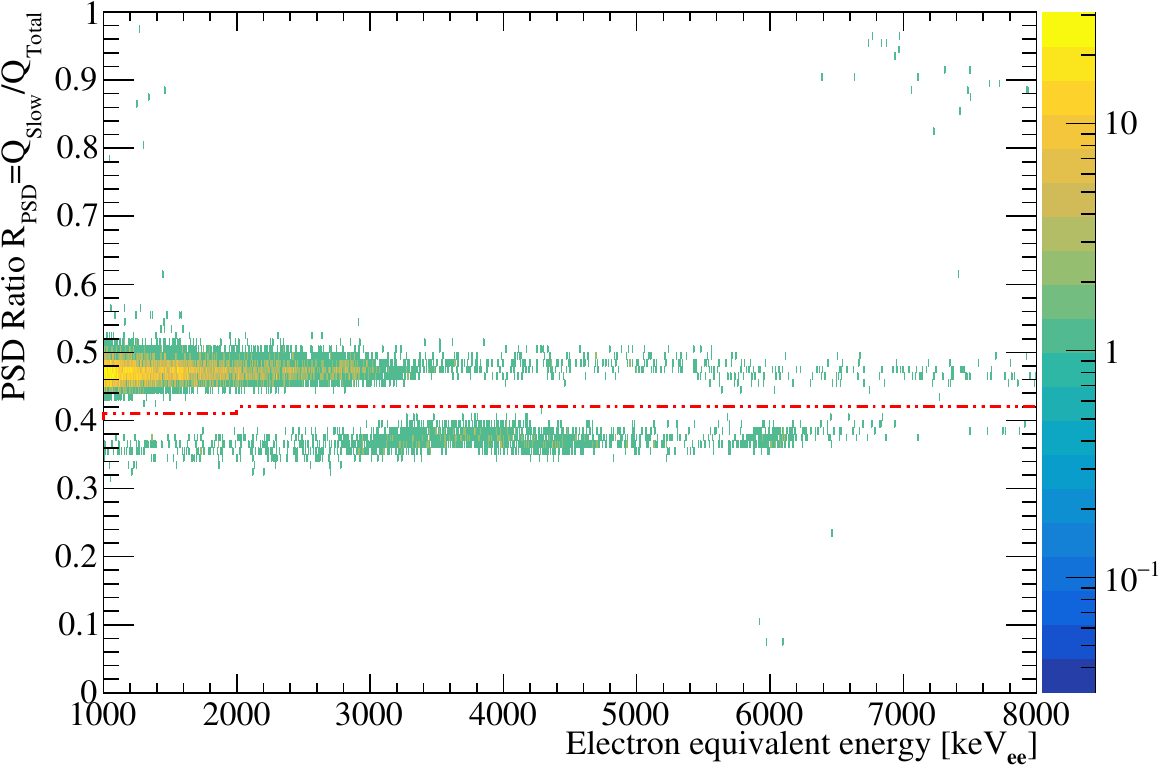}
  \caption{Ingot\#94 $R_\mathrm{PSD}$ result for the electron equivalent energy. Red dash line is \alpharay events extraction threshold.}
  \label{fig:energy-vs-rpsd}
\end{figure}
\begin{figure}[!h]
 \centering
 \includegraphics[width=0.65\columnwidth]{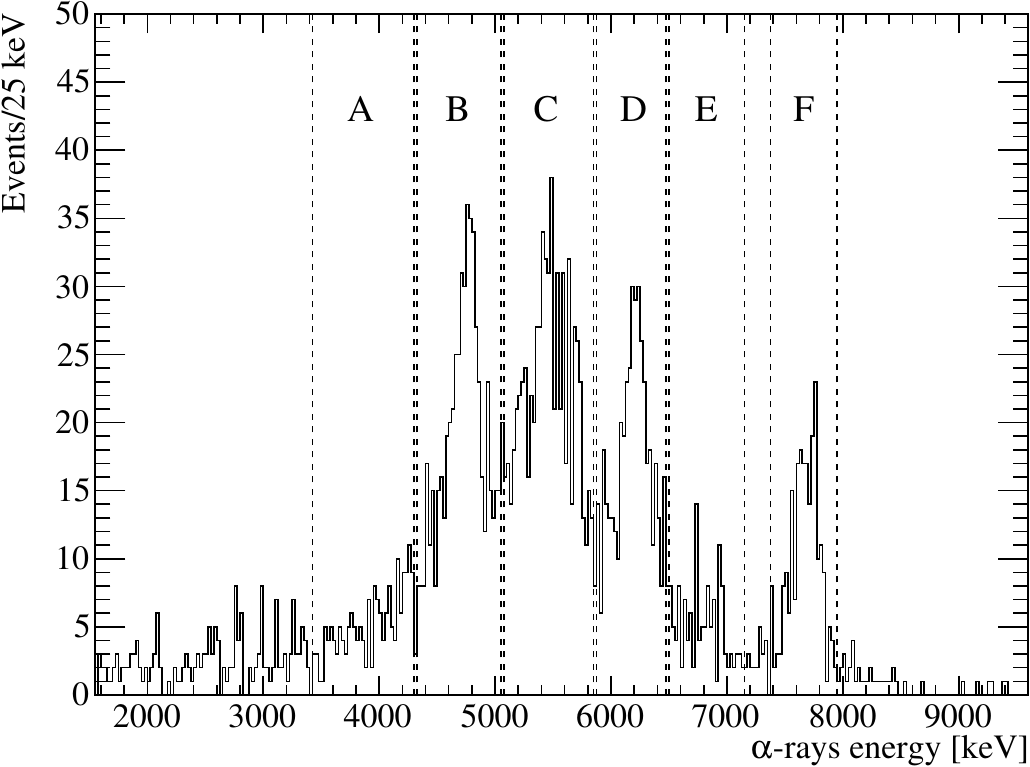}
 \caption{
$\alpha$-ray energy spectrum. The horizontal-axis represents the calibrated energy of alpha-ray emissions. Six peaks are observed, corresponding to RIs in the Th- and U-series. The number of events is estimated assuming the radioactive equilibrium.
}
 \label{fig:Alpha-raySpectZoom}
\end{figure}

The difference in pulse shapes allows us to discriminate \betagamma ray events from those induced by \alpharay events.
We applied Pulse Shape Discrimination (PSD) to distinguish events caused by \alpharay events from all other events.
The pulse-shape parameter named $R_\mathrm{PSD}$ is defined as a ratio of integrated currents over two different time windows after the pulse starts:
\begin{equation}
R_{\mathrm{PSD}} \equiv \dfrac{Q_{\mathrm{Slow}}}{Q_{\mathrm{Total}}}=\dfrac{\int_{200~\si{ns}}^{1200~\si{ns}}I(t) dt}{\int_{0~\si{ns}}^{1200~\si{ns}}I(t) dt},\
\end{equation}
where $I(t)$ is the current of the pulse. $Q_{\mathrm{Total}}$ and $Q_{\mathrm{Slow}}$ are the integrated charges over two different time intervals since the start of the pulse: [0~ns, 1200~ns] for $Q_{\mathrm{Total}}$, [200~ns, 1200~ns] for $Q_{\mathrm{Slow}}$.
The start time of pulse is determined by tracing backward in time from the maximum amplitude of waveform. The threshold is optimized using the single-photoelectron distribution.
A result of $R_\mathrm{PSD}$ for the electron equivalent energy is shown as a scatter plot in \figrefin{fig:energy-vs-rpsd}.
Due to a shorter time constant, the \alpharay events are distributed below the \betagamma rays in the energy region between 1000--8000~\keVee. The threshold for extracting \alpharay events was determined by the red dashed line that separates the $\alpha$ and \betagamma rays.
The energy spectrum of $\alpha$-rays is shown in \figrefin{fig:Alpha-raySpectZoom}. 
We identified six prominent peaks between 3000~\keVee \ and 8000~\keVee, which correspond to $\alpha$-rays from the U-series and Th-series. 
We analyzed each RI in the prominent peaks as shown in \tabref{Tab:RIConcentrationIngot94}, where the symbols in brackets denote the decay series.
Closely located peaks are combined into one peak.
For example, two peaks (index A in \tabref{Tab:RIConcentrationIngot94}) from \RI{238}{U} and \RI{232}{Th} cannot be resolved due to the limitation of energy resolution.
\begin{table}[!h]
 \caption{RI events with alpha-ray in Ingot\#94.}
 \label{Tab:RIConcentrationIngot94}
 \centering
 \begin{tabular}{cccc} \hline
 Index & RIs                         & Energy range [\si{keV}] & Events \\ \hline
 A & \RIU{238}{U},\ \RITh{232}{Th}                 & \vrange{3425}{4300}  & \vals{190}{20} \\
 B & \RIU{234}{U},\ \RIU{230}{Th},\ \RIU{226}{Ra}          & \vrange{4325}{5050}  & \vals{570}{30} \\
 C & \RITh{228}{Th},\ \RITh{224}{Ra}*,\ \RIU{222}{Rn},\ \RIU{210}{Po} & \vrange{5075}{5850}  & \vals{700}{30} \\
 D & \RIU{218}{Po},\ \RITh{212}{Bi},\ \RITh{224}{Ra}*,\ \RITh{220}{Rn} & \vrange{5875}{6475}  & \vals{430}{20} \\
 E & \RITh{216}{Po}                       & \vrange{6500}{7150}  & \vals{140}{20} \\
 F & \RIU{214}{Po}                       & \vrange{7375}{7925}  & \vals{230}{20} \\ \hline
 \end{tabular}
\end{table}
\tabref{Tab:RIConcentrationIngot94} represents a set of equations describing the number of events for each isotope. The numbers of \RI{232}{Th}, \RI{226}{Ra}, and \RI{210}{Po} events are calculated for uranium- and thorium-series isotopes under different assumptions of radioactive equilibrium.
All isotopes in the thorium series are assumed to be in radioactive equilibrium. 
For the uranium series, we consider the decay series $\RI{238}{U}\to\RI{234}{Th}\to\dots\to\RI{234}{U}$, and 
$\RI{226}{Ra}\to\dots\to\RI{214}{Pb}\to\dots\to\RI{214}{Po}\to\RI{210}{Pb}$. 
In the following, an arrow indicates a branching radioactive decay for each isotope. 
Nuclei with branching ratios below 1\% are neglected. % in the uranium series.
Taking into account the propagation of uncertainty, the number of RI events $N_\mathrm{RI}$ is obtained as $N_{\RI{232}{Th}}=N_{\RI{216}{Po}}= \vals{140}{20}$, $N_{\RI{226}{Ra}}=N_{\RI{214}{Po}}=\vals{230}{20}$ and $N_{\RI{210}{Po}}=\vals{280}{40}$. The RI concentration $C_{\mathrm{RI}}$ [\si{Bq/kg}] was calculated using the formula
\begin{equation}
C_{\mathrm{RI}} = \dfrac{N}{t M}\pm \dfrac{\sigma_N}{t M},
\end{equation}
where $N$ and $\sigma_N$ are number of RI events and statistical uncertainty, $t$ is lifetime and $M$ is the mass of NaI(Tl) crystal.
The exposure of Ingot\#94 $tM$ was \exposure. RI concentrations are shown in \tabref{Tab:RIConcentrationNewPICOLON}.
The BG stability of the NaI(Tl) crystals was confirmed, as no significant increase in concentration was observed concerning the BG results (see \refref{Kotera2023picolon}) for 30 days measured before the present experiment for Ingot\#94.
\begin{table}[!h]
 \caption{Concentration of RIs in Ingot\#94 for exposure of \exposure. Compared with the BG results of Ingot\#94 measured 30 days before the present experiment \cite{Kotera2023picolon}, no significant increase in concentration was observed, confirming the BG stability of the NaI(Tl) crystals.}
 \label{Tab:RIConcentrationNewPICOLON}
 \centering
 \begin{tabular}{ccccc} \hline
    RI& Ingot\#94 & \#94 First (30 d) result \cite{Kotera2023picolon} & Our goal & DAMA/LIBRA \cite{BERNABEI2008297}\\\hline
    $^{232}$Th [\activitykg] & \vals{7.4}{0.7} & \vals{4.6}{1.2} & $< 10$ & \vrange{2}{31}\\
    $^{226}$Ra [\activitykg] & \vals{12.6}{0.9} & \vals{7.9}{4.4} & $\sim 10$ & \vrange{8.7}{124}\\
    $^{210}$Pb [\activitykg] & \vals{15}{2}  & \vals{19}{6} & $< 50$ & \vrange{5}{30}\\\hline
  \end{tabular}
\end{table}
\section{Radioactivity sources external to NaI(Tl)} \label{sec:betagammaspect}
\begin{figure}[!h]
 \centering
 \includegraphics[width=0.65\columnwidth]{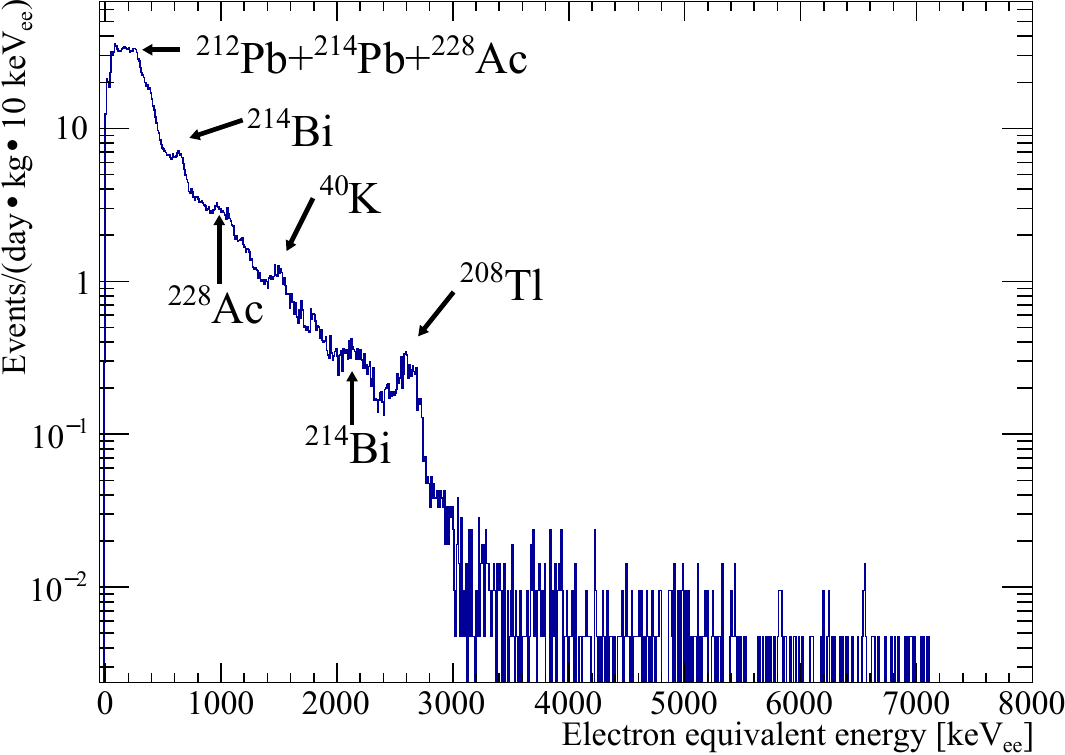} 
 \caption{
An excess peak is observed due to RI contamination from the external NaI(Tl) crystal. The PMTs and the oxygen-free copper shielding are considered the primary sources of the contamination.
}
\label{fig:betagammaspect}
\end{figure}
\figref{fig:betagammaspect} shows the \betagamma spectrum after applying the PSD cut for \alpharay rejection. We confirmed the presence of the excess peaks from radioisotopes external to NaI(Tl) crystals.
The amount of radioactivity in \tabref{Tab:RIConcentrationNewPICOLON} is at the same level as that of the PICOLON crystal previously reported in the literature \refref{Kotera2023picolon,fushimi2021development}.
The energy spectra shown in the \figrefin{fig:betagammaspect} exhibit event rates approximately ten times higher than those expected from surrounding the NaI(Tl) crystal, as estimated using the radioactivities listed in \tabref{Tab:RIConcentrationNewPICOLON}.
To identify the BG sources in \figref{fig:betagammaspect}, we measured the PMT(Hamamatsu R11065-20) using a high-purity germanium detector. As a result, multiple excess gamma-ray peaks were confirmed. The identified peaks are listed below.
The \RI{208}{Tl} (\genergy{2614.5}, \qbeta{5.0}, \halflive{3.04}{min}) and
\RI{40}{K} (\genergy{1460}, \qec{1.5}, \halflive{1.3\times10^9}{y}) peaks originate predominantly from material surrounding the NaI(Tl) crystals. The \RI{214}{Bi} (\genergy{2614.5}, \qbeta{3.3}, \halflive{19.4}{min}) peak contains events exceeding the internal BG amount estimated from inside of the NaI(Tl) crystal.
The remaining excess in the observed event rate is attributed to \RI{210}{Pb} contamination in the PMT circuit.
The \RI{228}{Ac} peak at \genergy{911} originates from radioactive contaminants in the components of the PMT divider circuit. In addition, the peak around \genergy{300} was identified as a single unresolved structure within the FWHM, attributed to the combined contributions of \RI{212}{Pb}, \RI{214}{Pb}, and \RI{228}{Ac}. 
\section{Noise reduction and detection efficiency} \label{sec:noisereduction}

In the low-energy region below \SI{100}{\keVee}. Noise events originate from Cherenkov radiation and the dark current of PMTs.
We applied four noise reduction techniques: (1) single-pulse rejection to separate noise from scintillation pulses, (2) baseline noise filtering to eliminate anomalous baseline fluctuations, (3) PSD-based rejection of noise events below 20~\keVee, and (4) \Qprehit reduction to reject anomalous charge pulses occurring before signals. 
The detection efficiency was estimated by applying all noise reduction techniques to \RI{60}{Co} calibration spectrum. 
It was then obtained by performing a maximum likelihood fit with the simulated \RI{60}{Co} spectrum.

\begin{figure}[!h]
\centering
\includegraphics[width=0.65\hsize]{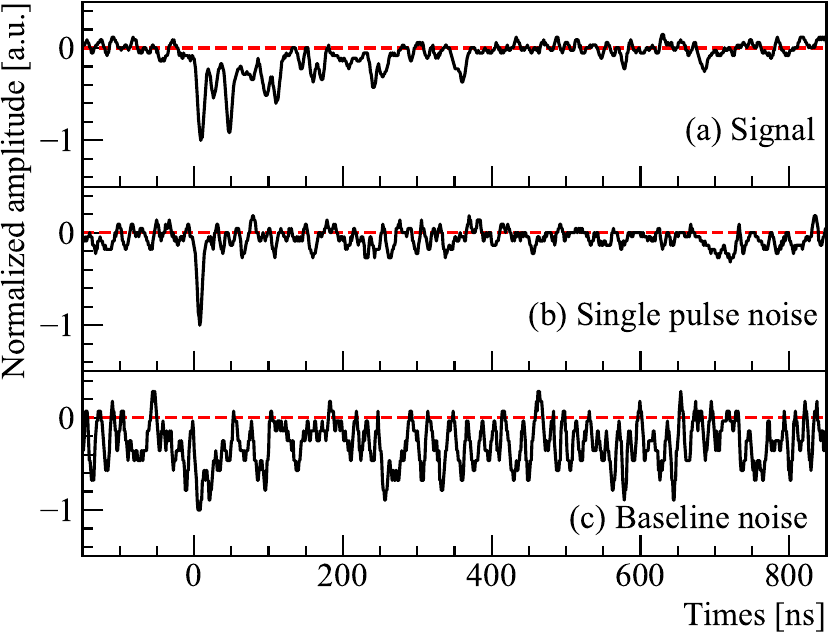}
\caption{Example of typical waveform at 4~\keVee\ signals. a vertical axis is a normalized amplitude of waveforms. (a) NaI(Tl) scintillation event signal. (b) single pulse noise event signal. (c) baseline noise event signal has $\Delta t>1000~\si{ns}$, where $\Delta t$ is the time difference between two single pulses from the start time of the primary pulse.}
\label{fig:Waveform_SignalandNise}
\end{figure}

Single pulse noise reduction is the primary noise rejection method.
\figref{fig:Waveform_SignalandNise}(a) and (b) shows typical scintillation and noise pulses.
The noise waveform consists of a single pulse followed by no pulse after \SI{200}{ns}.
The scintillation signal consists of many pulses spread over the scintillation decay time. Therefore, scintillation events are selected from all events using the timing window $\Delta t > \SI{200}{ns}$ after the first pulse,
where $\Delta t$ is the time difference between two single pulses, and the start time of a pulse is the same as the start time of the pulse in the PSD.
We defined a NaI(Tl) scintillation time constant (\SI{200}{ns}) hold-off window. When a subsequent pulse appeared within this window and its height rose above the threshold set at $\text{(Pulse maximum)}/e$, when $e$ is Euler's number. The effective width was redetermined as the interval from the start time of the pulse to the waveform time constant of the subsequent pulse. Repeating this rule over the sequence of pulses yielded the timing window $\Delta t$, corresponding to the time constant of the waveform.

\begin{figure}[!h]
\centering
\includegraphics[width=0.65\columnwidth]{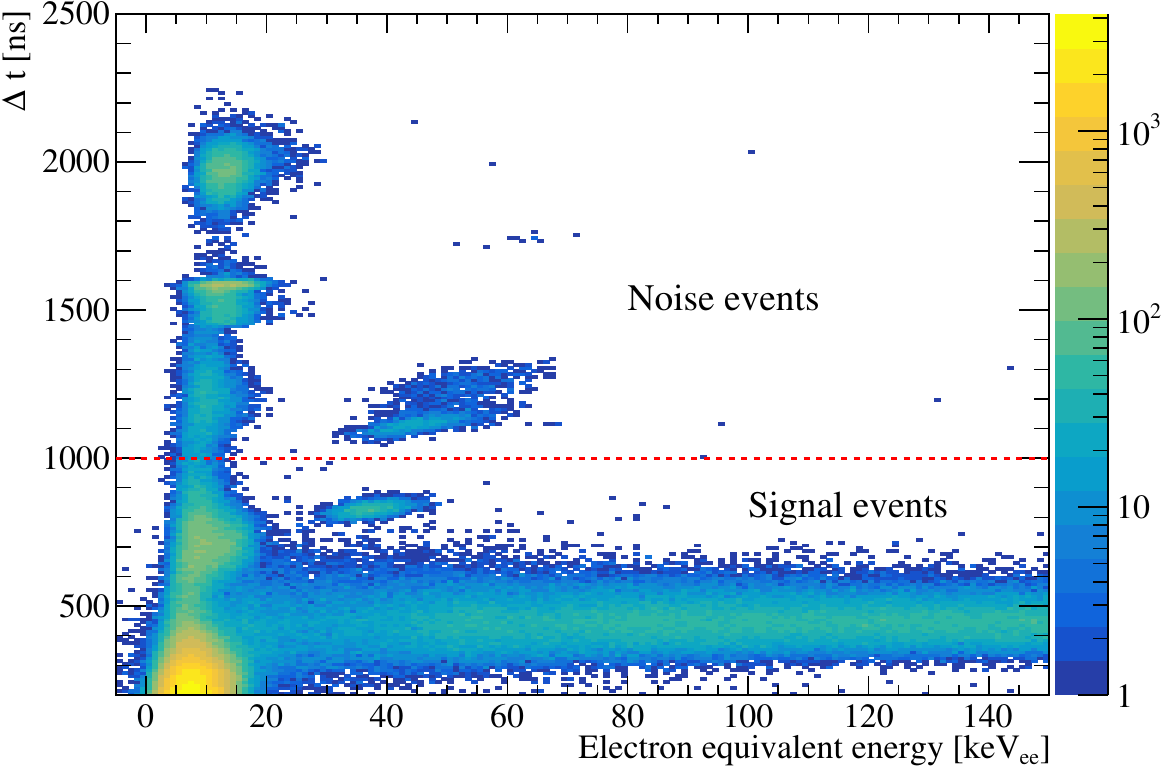}
\caption{The timing window parameter $\Delta t$ distribution for electron equivalent energy. Red dash line is primarily noise threshold. Remaining noise events were removed by other noise reduction methods.}
\label{fig:WidthAnalysis}
\end{figure}

Secondary noise reduction methods are used to suppress events caused by baseline noise. Noise from VME bus traffic and power supply lines causes unstable baseline, causing it to fluctuate or shift and reducing the efficiency of single noise pulse reduction.
An example of a baseline noise waveform is shown in \figrefin{fig:Waveform_SignalandNise}(c).
\figref{fig:WidthAnalysis} shows the timing window $\Delta t$ distribution for electron equivalent energy after applied the single noise reduction method. The baseline noise corresponds to the region with $\Delta t>1000~\si{ns}$ below 100~\si{keV}, which does not include the scintillation signal.
This is because the timing window of the baseline noise is longer than that estimated for the NaI(Tl) scintillation signals.
The two loci on \figrefin{fig:WidthAnalysis}, (1): $(30,70)~\keVee$ and $\Delta t: (1000, 1300)~\si{ns}$; and (2) $\Delta t$ around $800~\si{ns}$ and $(25, 50)~\keVee$; are also confirmed to be baseline noise as shown in \figrefin{fig:Waveform_SignalandNise}(c). These events were removed by the PSD noise reduction method shown next.
\begin{figure}[!h]
\centering
\includegraphics[width=\columnwidth]{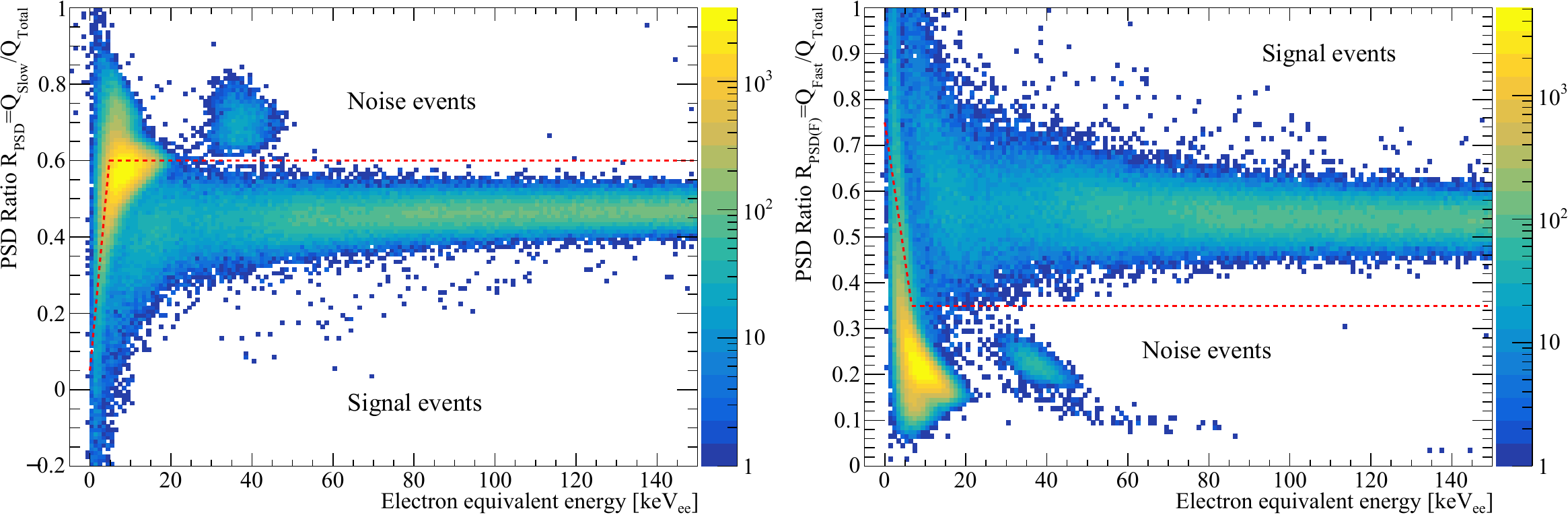}
\caption{Scatter plots of PSD ratio versus electron equivalent energy. Left: $R_\mathrm{PSD}$, right: $R_\mathrm{PSD(F)}$. The red dashed lines indicate the thresholds applied for double PSD noise rejection. Noise events that are classified as signal events in $R_\mathrm{PSD}$ are rejected by the threshold of $R_\mathrm{PSD(F)}$.}
\label{fig:PSDNoiseAnalysis}
\end{figure}

We applied double noise thresholds to the $R_\mathrm{PSD}$, $R_\mathrm{PSD(F)}$ versus the electron equivalent energy scatter plot in \figrefin{fig:PSDNoiseAnalysis} to reject PMT noise events below 80~\keVee. Where $R_{\mathrm{PSD(F)}}$ is a primary PSD parameter introduced for noise reduction, defined as
\begin{align}
R_{\mathrm{PSD(F)}} \equiv \dfrac{Q_{\mathrm{Fast}}}{Q_{\mathrm{Total}}}=\dfrac{\int_{0~\si{ns}}^{200~\si{ns}}I(t) dt}{\int_{0~\si{ns}}^{1200~\si{ns}}I(t) dt},
\end{align}
using the charge $Q_{\mathrm{Fast}}$ with a time window for [0~ns, 200~ns].
\figrefin{fig:PSDNoiseAnalysis}(left) shows the PSD distribution after applied the secondary baseline noise rejection method. \figrefin{fig:PSDNoiseAnalysis} (right) shows the event distribution for $R_\mathrm{PSD(F)}$ after applied the noise rejection to \figrefin{fig:PSDNoiseAnalysis} (left). A high-density distribution is noise events below \SI{100}{\keVee}. The dashed line indicates the threshold for noise rejection. All events removed by the cuts were noise clusters from PMT dark current. 
Baseline noise caused overlap of these single pulse noise signals.
Each baseline noise event had distinct amplitude and phase, and the waveforms overlap to form complex clusters.
These clusters complicated the PSD thresholds.

% 
% -- pre hit charge reduction
\begin{figure}[!h]
  \centering
  \includegraphics[width=0.65\hsize]{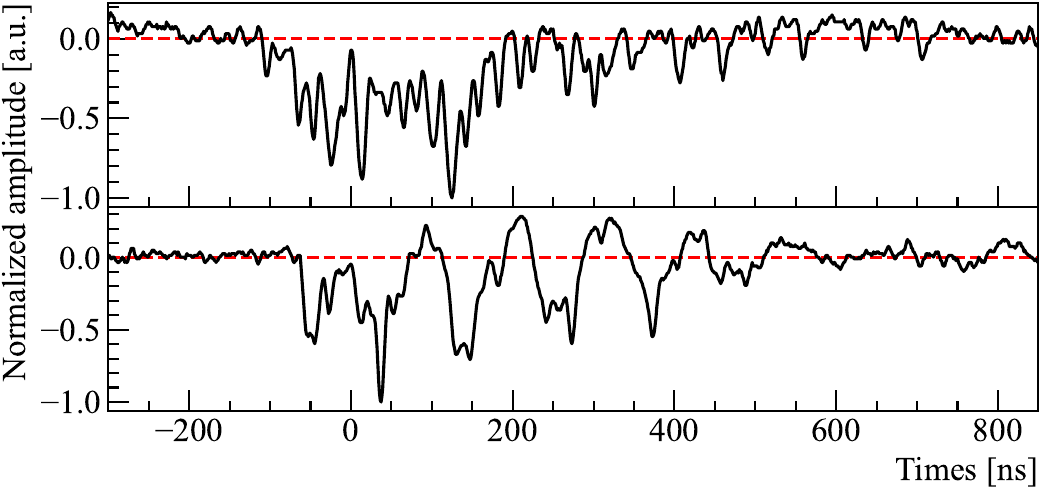}
  \caption{Example of typical wrong pulse waveforms bellow \SI{20}{\keVee}. Top: signal-like wrong pulse, bottom: noise-like wrong pulse.
  }
  \label{fig:Waveform_WrongPulse}
  \end{figure}

After 80~d from the start of the measurement on the realtime, we confirmed increased noise events bellow \SI{20}{\keVee}. 
% 原因
The trigger performance degradation was attributed to Ground instability accompanying the laboratory maintenance power outage immediately before \SI{80}{d}.
The example of the wrong pulse is shown in \figrefin{fig:Waveform_WrongPulse}. 
% 原因
From Ground instability, wrong trigger waveforms arise from ringing-like noise or from its overlap with scintillation waveforms.
To remove those events, We defined a pre-hit charge \Qprehit, as the integrated charge within \SI{100}{ns} prior to the start of pulse as

\begin{align}
  \Qprehit \equiv \int_{\SI{-100}{ns}}^{\SI{0}{ns}}I(t) dt.
\end{align}

\begin{figure}[h]
  \centering
  \includegraphics[width=0.65\hsize]{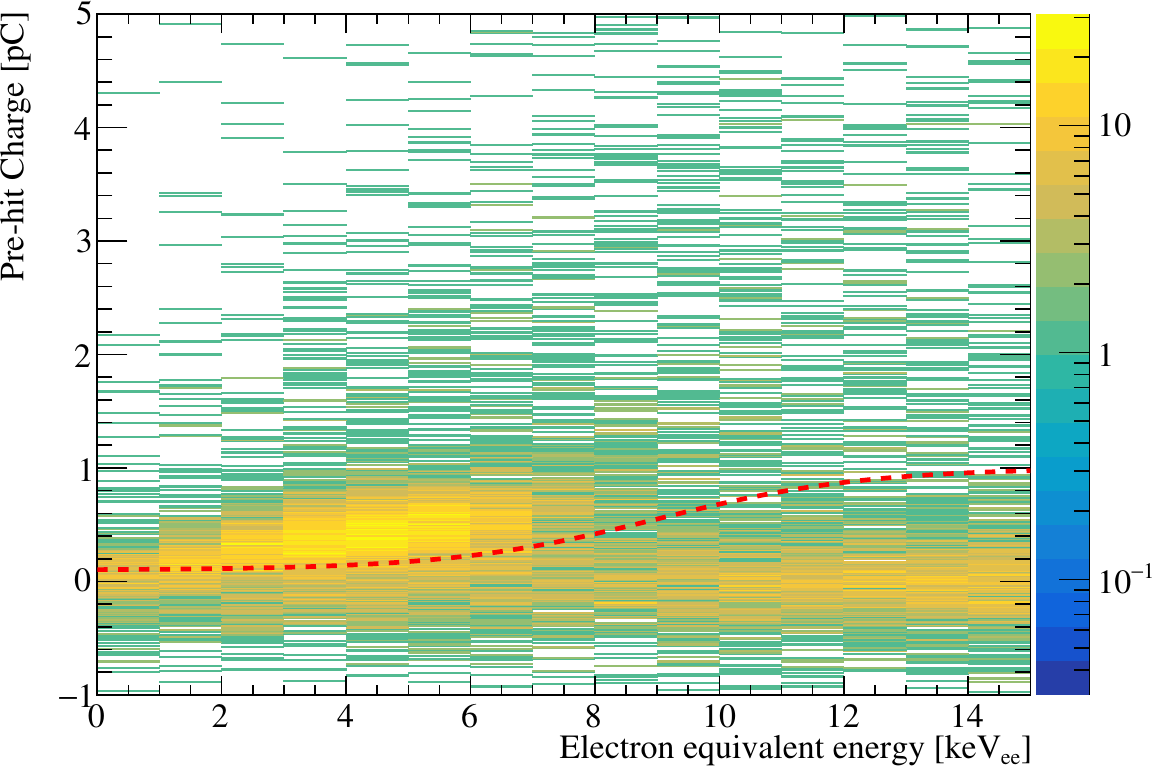}
  \caption{Scatter plot of \Qprehit versus electron equivalent energy. 
  Events with \Qprehit exceeding the threshold (red dashed line) include pulses caused by wrong trigger timing.
  }
  \label{fig:pre-hit-map}
\end{figure}

\figref{fig:pre-hit-map} shows a scatter plot of \Qprehit versus electron equivalent energy. For normal scintillation or noise pulses, \Qprehit distributes symmetrically around zero, reflecting baseline integration. 
In contrast, waveforms with wrong trigger timing show a excess in \Qprehit compared to the baseline distribution, allowing for event discrimination.
We applied a flexible rejection threshold, indicated by the red dashed line, to suppress the excess component that begins to overlap with the baseline region below \SI{10}{\keVee}.

\begin{figure}[h]
\centering
\includegraphics[width=0.65\hsize]{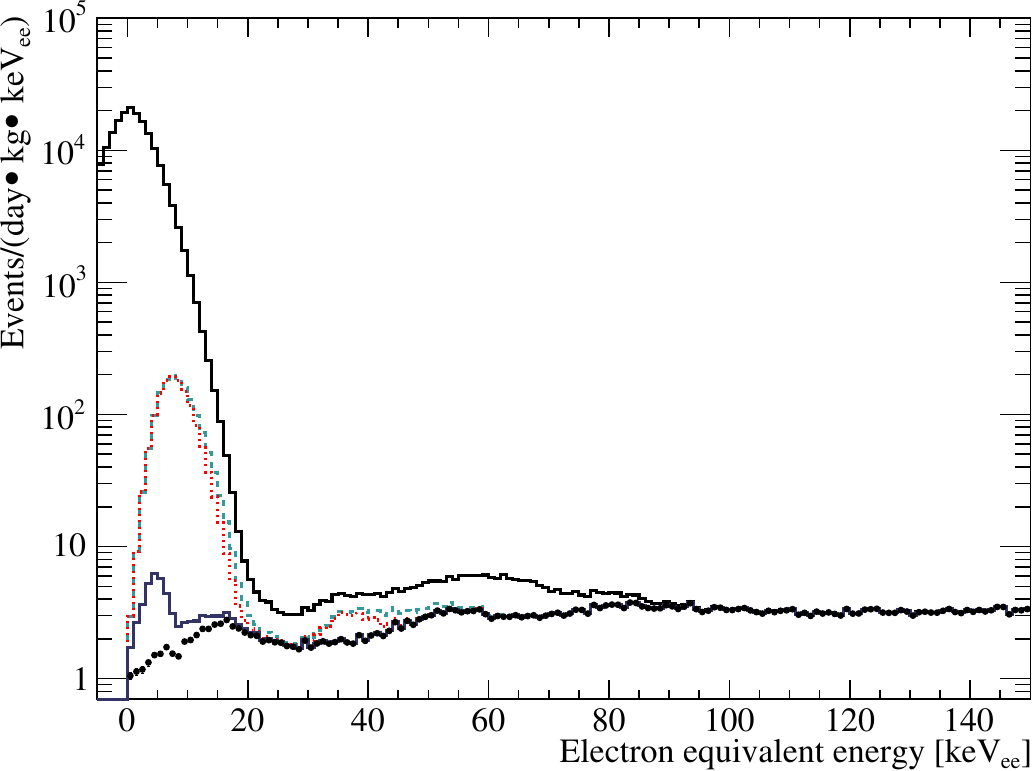}
\caption{Low energies energy spectra after applied noise reduction. The black line is no noise cutting. Green dashed line: after the single noise reduction method. Red dashed line: before applied double PSD cutting. blue line: charge rejection using \Qprehit method. The final result shows the black point.}
\label{fig:noisereduction-result}
\end{figure}

% -- result
\figref{fig:noisereduction-result} shows the energy spectrum before and after applied all reduction methods. A significant suppression of noise events is observed after the cuts.
\begin{figure}[h]
  \centering
  \includegraphics[width=0.65\hsize]{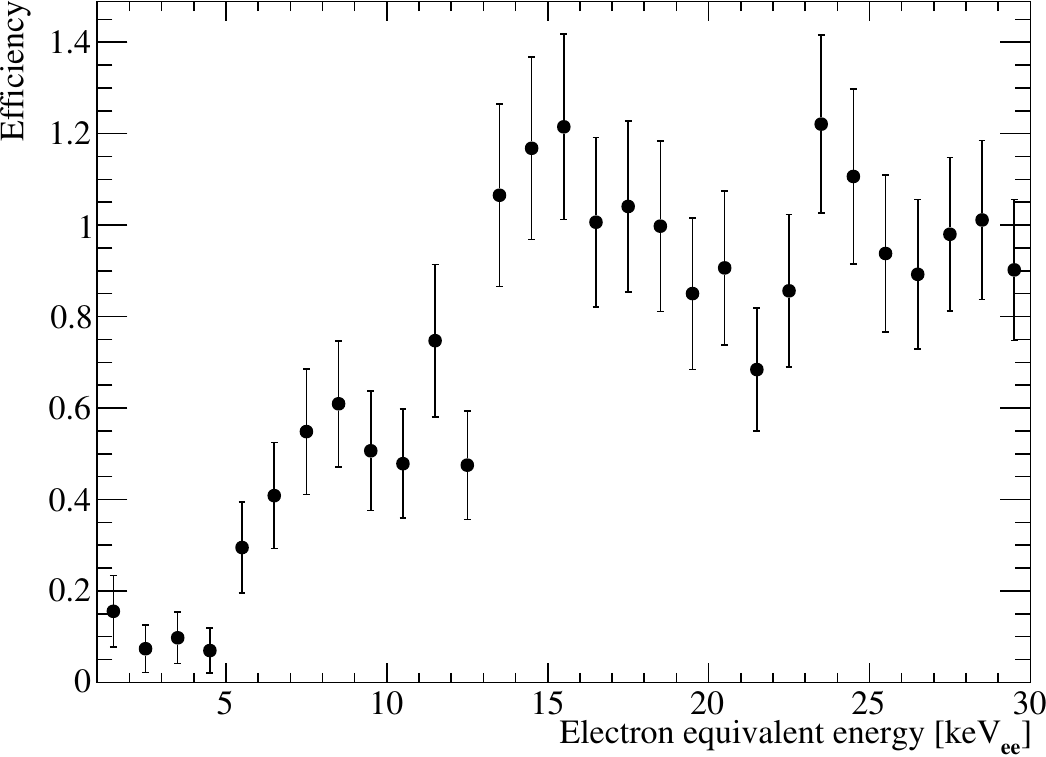}
  \caption{The detection efficiency, determined by fitting the simulated \RI{60}{Co} spectrum to the measured calibration spectrum.}
  \label{fig:detector-efficiency}
 \end{figure}
The efficiency of noise reduction is as follows: (1) Single pulse reduction for all events: 99.1\%; (2) baseline noise filtering for single pulse reduction: 7.8\%; (3) double PSD noise reduction method after baseline noise filtering method: 85.4\%; (4) applied \Qprehit charge rejection method: 12.8\% after applied double PSD noise reduction method.
The overall noise selection inefficiency was less than 0.1\%.
\figref{fig:detector-efficiency} shows a detection efficiency for the energy below 30~\si{keV} simulated using the Geant4 Monte Carlo toolkit (version 11.1.1), which includes the G4RadioactiveDecay and G4EmLivermorePhysics process, following the analysis procedure adopted by \refref{adhikari_understanding_2017,adhikari_background_2018}.
% result
The significant uncertainty in the detection efficiency arises from the limited number of \RI{60}{Co} calibration data samples.
It was difficult to accumulate a sufficient number of source events due to the DAQ system instability under high-rate noise conditions.
The detection efficiency was estimated by scaling the simulated \RI{60}{Co} spectrum to the \RI{60}{Co} calibration spectrum using the likelihood method. 
The calibration spectrum was applied all noise reduction techniques. 
% added
Using the \RI{60}{Co} spectrum to estimate the detection efficiency is justified for the following reasons:
(i) Dark matter nuclear recoils are observed in electron equivalent energy due to quenching. 
Applying the sodium and iodide quenching factors suffices to evaluate the nuclear recoil spectrum in electron equivalent energy. 
(ii) The nuclear recoil spectrum of NaI(Tl) has a complex structure. 
In contrast, the \RI{60}{Co} spectrum has a flat structure in the low-energy region, which reduces the uncertainty in the detection efficiency and renders it suitable for evaluation.

\section{Limit on WINPs} \label{sec:Limit}
\begin{figure}[h]
\centering
\includegraphics[width=0.7\columnwidth]{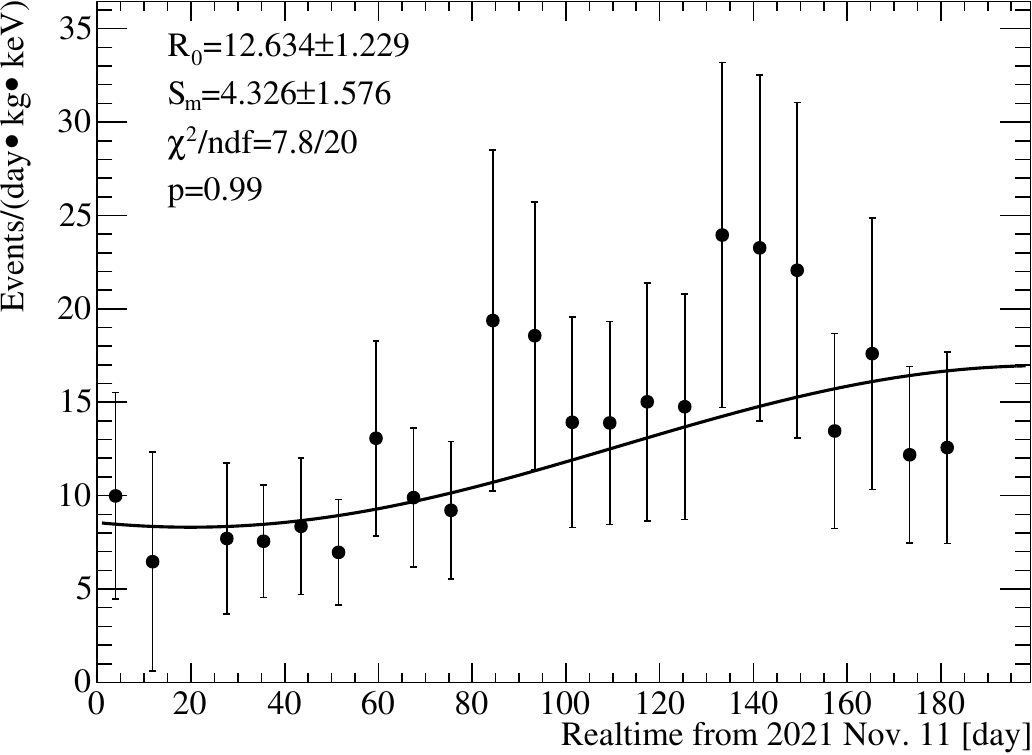}
\caption{The BG rate over realtime for \roirange ROI, since measurements stared on 2021 Nov. 11. Black line and text is a fitting result using the annual modulation function eq.~\eqref{eq:event-rate}. The finite value of $S_m$ is attributed to excess events in the latter noisy period and to uncertainties in the detection efficiency.}
\label{fig:BGRateROI}
\end{figure}

The limits on WIMPs were estimated by analyzing the annual modulation in the BG spectrum taken for six-month. A detailed description of the calculation method is given at the \refref{LEWIN199687}.
The annual modulation of event rate WIMPs is caused by changes in the relative velocity $\boldsymbol{v}_\mathrm{E}$ of the Earth and the WIMPs in the Solar system. The differential scattering rate is given by
\begin{align}
\dfrac{dR}{dE_\mathrm{R}}=\dfrac{N\rho_0}{m_\mathrm{\chi}}\int_{v_\mathrm{min}}^{v_\mathrm{max}} vf(\boldsymbol{v}+\boldsymbol{v}_\mathrm{E}) \dfrac{d\sigma}{dE_\mathrm{R}} d^3\boldsymbol{v}.
\end{align}
Where $E_\mathrm{R}$ is the kinetic energy of recoiled nucleus, $N$ is the number of target nuclei, $\rho_0,\ m_\mathrm{\chi}$ are local DM density ($0.4~\si{GeVc^{-2}cm^{-3}}$) and DM mass, $f(\boldsymbol{v}+\boldsymbol{v}_\mathrm{E})$ is the velocity distribution in the Earth's frame, $\sigma$ is the scattering cross section on the nucleus for WIMPs. The range of integration is defined as $v_\mathrm{min}\equiv (E_\mathrm{R}/E_0r)$ and $v_\mathrm{max}\equiv 2\mu v/m$, using WIMPs kinetic energy $E_0$ on WIMPs average velocity $v_0=230~\si{km/s}$, $m$ is mass of target nuclei,
$r\equiv 4m_\mathrm{\chi}m/(m_\mathrm{\chi}+m)^2$, $\mu$ is DM-nuclei reduce mass.
Taking into account the Earth's orbital motion and the Sun's peculiar motion, \si{12~km/s} for the direction of motion of the Solar system, the Earth WIMP's relative velocity varies as follows
\begin{align}
v_E=(v_0+12)+15 \cos \left\{ \frac{2\pi}{365.25~\si{d}}(t-t_0)\right\}~\si{km\ s^{-1}},
\end{align}
$t$ is a time of date. 
The velocity of $\SI{15}{km/s}=v_\mathrm{E}\cos \theta$ accounts for the $\theta=60^\circ$ inclination of the orbital plane relative to Earth's orbital motion at $v_\mathrm{E}=\SI{30}{km/s}$. The initial phase $t_0$ corresponds to June 2. Annual modulation causes a variation of approximately 6\% in $v_0$. A event rate $R$ is expressed as Taylor series:
\begin{align}
R = R_0 + S_\mathrm{m} \cos \left\{ \frac{2\pi}{365.25~\si{d}}(t-t_0)\right\}~\dru. \label{eq:event-rate}
\end{align}
$R_0$ is the non-modulating parameter. 
The amplitude $S_m$ is observed in the region below a few \si{keV} (range of interest: ROI), which corresponds to the WIMP-nucleus recoil energy. 
DAMA/LIBRA experiment report the $S_\mathrm{m}=0.0102\pm 0.0008~\dru$ \cite{10.21468/SciPostPhysProc.12.025}.
\figref{fig:BGRateROI} shows the time variation of the event rate in the \roirange ROI after applying the detection efficiency. 
The fitting function from the eq.\eqref{eq:event-rate} was applied. Generally, $t_0$ is set to $\SI{162.5}{d}$ based on the virial model of seasonal variation starting from 2nd June. However, in this study, the horizontal-axis of the graph is the time elapsed since the start of the measurement, so the phase difference from 2nd June was corrected and fitted using the maximum likelihood method.
\tabref{Tab:SmResult} shows the $S_m$ result of DAMA/LIBRA and other NaI(Tl)-based DM search groups and our result.
$S_m$ takes a finite value, which is considered to originate from excess events in the latter noisy period and from uncertainties in the detection efficiency.

The differential event rate needs to consider the quenching factor $f$ of the NaI scintillator.
For sodium nuclei, $f_\mathrm{Na}=0.173\pm0.001$ from neutron beam experiments on PICOLON NaI(Tl) crystals by \refref{Urano:2024/M}. For iodine, $f_\mathrm{I}=0.051\pm0.002$ from \refref{JOO201950}.
In the present analysis, it is assumed that there is no energy dependence on the quenching factor.
The limit on dark matter (DM) interactions is estimated using the experimentally obtained amplitude $S_m$ and the theoretical expectation $S_m^0$.  
The value of $S_m^0$ is derived from the difference in the event rate of the energy spectrum within the \roirange ROI, as predicted by theoretical calculations.
Theoretically, the spin-independent (SI) WIMP-nucleon cross-section $\sigma^0$ was computed assuming equal couplings for protons and neutrons. Based on these parameters, the limit on the experimental cross-section $\sigma$ can be determined using the ratio $S_{\mathrm{m}} : S_{\mathrm{m}}^0 = \sigma : \sigma^0$.

\begin{table}[!h]
  \caption{$S_m$ result by PICOLON and other NaI(Tl)-base DM search groups.}
  \label{Tab:SmResult}
  \centering
  \begin{tabular}{ccc} \hline
    ROI [\si{keV_{ee}}] & Group  & $S_m$ [\dru] \\\hline
    2--6 & DAMA/LIBRA \cite{10.21468/SciPostPhysProc.12.025} & $0.0102\pm0.0008$\\
         & ANAIS \cite{ANAIS2024Modulation} & $0.0031\pm0.0037$\\
         & COSINE \cite{carlin2024cosine100datasetchallengesannual}    & $0.0053\pm0.0031$\\
         & PICOLON & $4.3\pm1.6$\\\hline
  \end{tabular}
 \end{table}

\begin{figure}[h]
\centering
\includegraphics[width=0.7\columnwidth]{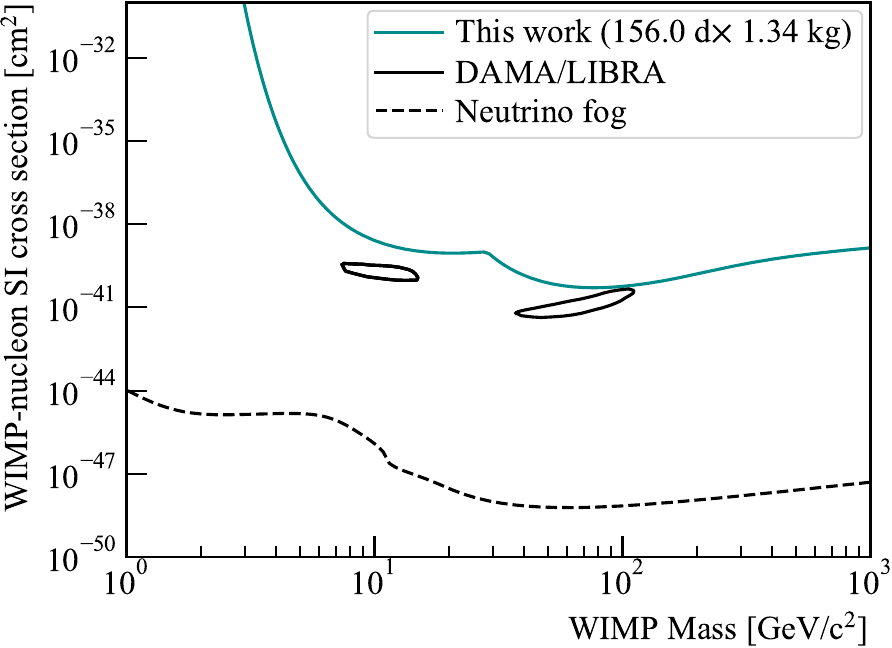}
\caption{The DM--nucleon SI limits for \roirange ROI \cite{10.1093/ptep/ptac097,XENON1T2016,xenon2023dark, PhysRevLett.127.251802}. The green line is the limit of the present work, \DMupperlimit (see \tabref{Tab:SmResult}). }
\label{fig:Limit}
\end{figure}

\figref{fig:Limit} shows the WIMP--nucleon SI limit interaction.
The black line is the DAMA/LIBRA search area, and the green line is the limit calculated when the analysis excludes noise sections. PICOLON NaI(Tl) crystal, Ingot\#94, have $\sim O(10^{-41})~\si{cm^2}$ cross section with \roirange ROI at \SI{100}{GeV/c^2} WIMP Mass.

\section{Summary and Conclusions}
This study presents the first six-month annual modulation result from the PICOLON experiment.
The measurement was conducted using highest-purity NaI(Tl) crystals at an independent experimental site, separate from other NaI(Tl)-based dark matter search experiment.
The current background rate is approximately 1.5~\dru on \roirange ROI.
The amplitude of the annual modulation takes a finite value. 
However, the observed effect arises from non-physical causes. 
The application of the \Qprehit noise reduction to suppress noise in the latter part of the measurement reduced the number of events and the detection efficiency.

Despite a small NaI(Tl) crystal mass, \SI{1.34}{kg}, the achieved sensitivity slightly approached the DAMA/LIBRA region under 
high-rate noise conditions. 
The DAMA/LIBRA group has stated that adequate verification requires NaI(Tl) crystals with radiopurity comparable to those used in their experiment~\refref{10.21468/SciPostPhysProc.12.025}. 
Contamination from \RI{210}{Pb} and \RI{40}{K} in the crystal causes large fluctuations in the event rate.
PICOLON purest NaI(Tl) crystal exhibited background stability without time dependence. 
PICOLON is the only experiment that has demonstrated stable production of ultra-pure NaI(Tl) crystals with such high radiopurity. 
 Therefore, overcoming the remaining challenges will enable adequate verification of the DAMA/LIBRA signal.

Development is in progress on a double-readout detector and a machine-learning-based noise reduction algorithm to suppress noise from the PMTs and DAQ system.
The dual readout detector is expected to effectively suppress noise and reduce the need for complex software cuts.  
The DAQ system has been upgraded. Due to the improved noise tolerance and stabilized data acquisition, sufficient statistics for detector efficiency and background estimation are now being accumulated.  
A machine-learning-based noise reduction method is currently in progress, and its effectiveness is being evaluated using a new dataset.
These developments are expected to further improve both noise reduction and detection efficiency, leading to a reduction in the uncertainty of event rates within the range of interest.
\section*{Acknowledgment}
This work was supported by JSPS KAKENHI (Grant Nos.: 26104008, 19H00688, and 20H05246), JST SPRING (Grant No. JPMJSP2113), a discretionary expense of the president of Tokushima University, and the World Premier International Research Center Initiative (WPI Initiative). The KAMIOKA MINING \& SMELTING Co. provided service for activities in the mine.

\bibliographystyle{ptephy}
\bibliography{Author_tex}
\end{document}